# Computer-Assisted Processing of Intertextuality in Ancient Languages

**Mark Hedges[1*], Anna Jordanous[2], K. Faith Lawrence[1], Charlotte Roueché[1], Charlotte Tupman[3]**

1 King's College London, UK

2 University of Kent, UK

3 University of Exeter, UK

*Corresponding author: Mark Hedges mark.hedges@kcl.ac.uk

**Abstract**
The production of digital critical editions of texts using TEI is now a widely-adopted procedure within digital humanities. The work described in this paper extends this approach to the publication of gnomologia (anthologies of wise sayings), which formed a widespread literary genre in many cultures of the medieval Mediterranean. These texts are challenging because they were rarely copied straightforwardly; rather, sayings were selected, reorganised, modified or re-attributed between manuscripts, resulting in a highly interconnected corpus for which a standard approach to digital publication is insufficient. Focusing on Greek and Arabic collections, we address this challenge using semantic web techniques to create an ecosystem of texts, relationships and annotations, and consider a new model – organic, collaborative, interconnected, and open-ended – of what constitutes an edition. This semantic web-based approach allows scholars to add their own materials and annotations to the network of information and to explore the conceptual networks that arise from these interconnected sayings.

## INTRODUCTION
The TEI (Text Encoding Initiative) XML format has been widely adopted as the standard encoding for marking up textual data with semantic content [Mylonas & Renear, 1999; Pierazzo, 2011; Sperberg-McQueen, 1991]. The adoption of this standard in principle facilitates interoperability between different resources, enabling them to be used in combination for new research, and this publication strategy has been embraced widely in the digital humanities community.[i] Lack of communication and failure to share research can still result in consolidation rather than expansion of information, the so-called 'digital silo' [Nichols, 2009; Zorich, 2008], but sometimes we also need to be able to do more with our texts than TEI currently allows.

We are therefore extending the TEI model through our work on editing medieval *gnomologia*. It has long been realised that philosophical, moral and scientific ideas have travelled, both within and beyond their own cultures, not only through the transmission of complete texts, but in collections of citations and summaries. These collections survive in abundant medieval manuscripts, which are not very rewarding to publish, and it is not easy to illustrate such processes within the confines of print. The Sharing Ancient Wisdoms project (SAWS), funded by HERA from 2010 to 2013, aimed to analyse some of the collections known as *gnomologia*: collections of wise sayings containing moral or social advice, or expressing philosophical ideas. [ii] Such collections of extracts from earlier works were rarely straightforward copies; sayings were selected from other manuscripts, reorganised, and modified or reattributed. They also crossed linguistic barriers, e.g. from Greek into Arabic, again rarely in straightforward translation; changes often reflected a change of social context, especially between different cultural traditions. In later centuries, collections were translated from Arabic into western European languages.[iii] In all languages, such collections also informed the writing of continuous texts, of a kind that are more readily perceived as literary. The project aimed to examine such texts, publishing several gnomologia in Greek, Arabic and Latin, as well as a number of continuous texts that used gnomologia as sources. These complex traditions themselves call into question the simple concept of citation: many examples of familiar passages may come not from full copies of the original text, but through a chain of texts, which may extend over centuries. Moreover, the compilation of collections, which to the modern reader appears a second order activity, required a creative process; each collection was shaped in some way, and this must be taken into account if we are seeking to understand the preoccupations and concerns of particular periods. While we could not hope to present more than a fraction of this rich and abundant material, we aimed to develop tools and protocols for doing so, in order to reveal, and analyse, some of the transitions between texts.

In this paper we describe a methodology and framework for publishing gnomological manuscripts that addresses and exploits their high degree of connectivity, without imposing a false concept of hierarchy. The paper considers the rationale for the work in Section 1. After a survey of related research in Section 2, in Section 3 we describe our information model, including a brief description of the ontology. We describe our approach to implementation in Section 4; our evaluation and future work in Section 5; and our current conclusions in Section 6.

## I GNOMOLOGIA AND DIGITAL EDITIONS

There has long been interest in the relationships within and between these manuscripts [Gutas, 1981; Richard, 1962; Rodríguez Adrados, 2001], as the analysis of these interrelations can reveal much about the dynamics of the cultures that created and used these texts. The large number of manuscripts, the complexity of their interrelationships, and the fact that a certain critical mass of material is required to carry out such research, has hindered their exploitation in the past; these very factors, however, make research in this area particularly susceptible to digital methods.

The nature of the material suggests a fundamentally different approach to creating an edition. Instead of considering variant witnesses to an 'original' text[iv] of which we are trying to create a single edited version, we have a number of interrelated texts of equal standing. The similarities between these texts need to be represented, but as these similarities take various forms, have various degrees, and operate at various levels of granularity, a more nuanced approach is required.

We are extending rather than rejecting the standard paradigm. We envisage that any particular text will be edited using TEI, and that such editions will continue to be published in digital libraries. We are, however, concerned not *only* with creating digital editions of these texts; we are building on current best practice, to publish the gnomologia in a manner that enables a better understanding of these texts as a network of information rather than as isolated documents. As long ago as 1990, DeRose and his co-authors reflected on how electronic text documents could best be structured for flexibility in use and reuse.[v] We are addressing this by better facilitating linking and comparisons, using an approach based on semantic web technologies.

The texts that we are editing form a small subset of these manuscripts, the majority of which will be edited and published by others; we thus envisage scenarios in which other groups will link their texts to ours (and to those of still other scholars). Many will identify relationships between texts that they publish independently as semantic triples. Consequently we are creating a framework of tools and methods that will enable researchers to add texts and relationships of their own, to create a corpus whose value will increase with its size and interconnectivity. We envisage an eventual network of marked-up texts and textual excerpts, linked together to allow researchers to represent, identify and analyse the flow of knowledge and transmission of ideas through time and across cultures.

This will enable scholars to create a more detailed picture of anthological sources, and provides a clearer picture of what was read and deemed important at a particular time and place. The extensibility of our approach means that others will be able to link the remaining unidentified text passages to sources they have identified, thus building and strengthening the corpus.

This publishing model also serves to contribute to the wider debate around the somewhat ambiguous notion of citation. Traditional editions can often indicate in a broad-brush manner that one section of text is a citation of another earlier text, without a deeper discussion of the nuances of the term, or without questioning the author's access to the earlier works. This model will enable us to understand better how often Byzantine - and indeed many medieval authors - use citations from collections, and how these collections may influence the thinking of an author. Some continuous texts may in fact be driven by the shape and selection of the collection with which they worked. It also enables us to address the question of what exactly constituted a 'learned' author in medieval Byzantium. This leads to a new concept of what constitutes an edition – a corpus that is organic, collaborative, interconnected, and open-ended.

While our work has focused on gnomologia, and in particular on Greek and Arabic gnomologia from the ninth to twelfth centuries AD, the methods and tools developed are applicable to other groups of manuscripts with analogous characteristics, such as medieval mathematical, medical or scientific texts.

## II RELATED WORK

A key aspect of SAWS is to represent relationships between and within collections of gnomic sayings. The RDF[vi] (Resource Description Framework) format is appropriate for this purpose, particularly when supported by an ontology of relevant information. We want to use RDF-like syntax to mark up relations between the text and links to external entities, and while RDFa (RDF with annotations) allows RDF to be encoded directly in marked-up documents, it has primarily been deployed in XHTML documents. It would be desirable to extend the scope of RDF to a wider scale [Eide et al., 2008][vii] to TEI XML documents, without extensive changes being required to the XML or to the workflow. This last point is of particular concern for the growing community of non-technical users of TEI [Pierazzo, 2011]. Keeping structural, syntactic and semantic information in the same documents also makes the markup process simpler and potentially less error-prone. To date, no method for accommodating TEI and RDF in the same document has been adopted as standard by the TEI community, though several approaches have recently been offered.

RDFTEF [Tummarello, 2005] is a Java-based tool for converting TEI files to a form which can incorporate and output RDF/XML markup. It implements a basic ontology for representing structural and syntactical elements and allows additional ontologies to be added. Queries need to be relatively complex and standard XML tools cannot be deployed within the RDFTEF environment [Portier et al., 2012]. RDFTEF has been criticised as '[o]nly a "toy" experiment' [Portier et al., 2012] for these limitations and lack of ongoing maintenance (last source code update 2007). Also, RDFTEF introduces a new stage of work to the existing editing workflow and requires extra software.

RDFa has been used to encode RDF in a TEI document [Jewell, 2010; Lawrence 2011]. This work primarily used the OntoMedia (OM) ontology [Lawrence, 2007][viii] to describe elements within the textual narrative and to annotate the TEI XML with explicit reference to the ontological class of the typed event or entity. This typing was on an automatic basis, processing information extracted from the TEI via a conceptual mapping between the TEI and OM. A second script was used to generate RDF linked data from the extended TEI. By drawing on the ontological data held in the RDFa as well as the information in the structure and elements of the TEI, triples were created that could be cross-referenced internally and to external data resources while retaining a link back to their textual context. Specialised scripts had to be deployed to extract the RDF to add to a triple store. Deploying such scripts is non-trivial for non-technical users, in setting up the appropriate environment and in executing the scripts. The scripts used by Jewell's and Lawrence's work were also highly specific to those documents. These issues were also seen in a similar script-based approach to automated creation of RDF triples from TEI documents, in work performed by the Supporting Productive QueRies (SPQR) project [Blanke et al., 2012]. There is a more user-friendly alternative of transformations through XSLT stylesheets, incorporated into the user interface of tools like the Oxygen XML editor. Another tool is available to represent document structure(s) with RDF: the EARMARK OWL ontology [Peroni and Vitali, 2009]. This uses RDF to model structural information, but does not model the text and additional semantic information, so again structure, data and markup become separated. To reduce over-specificity and encourage re-use of our materials, the adoption of a more generic underlying model for transformations is explored in this present paper.

The inclusion of RDF in TEI documents is a current area of interest in the TEI-Ontologies Special Interest Group (SIG),[ix] which is using XSLTs to convert TEI to RDF [Ore and Eide, 2009] by relating TEI markup to vocabulary in the CIDOC-CRM cultural heritage model [Doerr, 2003]. CIDOC-CRM is the Cultural Reference Model for the museum heritage organisation CIDOC. The inclusion of FRBRoo, the Functional Requirements for Bibliographic Records (FRBR) model harmonised with CIDOC-CRM [Doerr and LeBoeuf, 2009], has also been discussed. However work in this area is progressing slowly and development has concentrated around TEI/CIDOC-CRM harmonisation, for example see [Ciula and Eide, 2014]. Some mappings have been drafted (last updated 2007/8) and stylesheets (last updated 2011) and guidelines (last updated 2010) have been published, all by the SIG, but two issues are worth noting:

- The size of the current TEI P5 tagset raises practical difficulties in providing comprehensive mapping from TEI to alternative representations. The TEI ontologies SIG has identified a subset of elements to map to CIDOC-CRM, choosing only those which represent semantically meaningful elements within the text, "such as persons, places, dates and events"[x]. This approach is practical but disregards many triples of potential interest such as document structure and metadata.

- The only direct representation of lexical material within CIDOC-CRM is through one class (E33 Linguistic Object) and its two subclasses (E34 Inscription, E35 Title). This choice of CIDOC-CRM as base model is acknowledged to be influenced by the research interests of the SIG members in cultural heritage and museum documentation[xi]. For our interests in structural information and metadata, the Dublin Core (DC) model seems a more natural choice and is highly developed and widely adopted. A mapping from TEI to DC has been tackled in stylesheets created by the SIG but does not appear in reports of their main approach.

The aim of SAWS is to represent semantic relationships between passages of text identified during the editorial process; it was consequently desirable to mark these up directly in TEI. As TEI did not provide a standard mechanism for encoding all such relationships, we proposed the adoption of the **<relation>** element, which was subsequently accepted by the TEI, for encoding RDF relations in a TEI document[xii], representing the Subject-Predicate-Object triple format of RDF through the following attributes: **@active**, **@ref** and **@passive**. We describe and explain our encoding of RDF relations using **<relation>** in Section 4.2.

Related work has also been carried out in the use of ontologies to represent information in manuscripts and cultural heritage objects; this is discussed in Section 3.1.

## III CONCEPTUAL MODEL

In this section we describe our conceptual model, addressing:

- The base ontology and the motivations for its selection.
- The extensions added to form the SAWS ontology.

### 3.1 Base Ontology

Our ontology reuses the FRBRoo ontology [Doerr & LeBoeuf, 2007], a combination of the CIDOC-CRM and FRBR ontologies. The CIDOC Conceptual Reference Model (CRM)[xiii] is an ontology of the information and relationships relevant for cultural heritage documentation [Doerr, 2003]. CIDOC-CRM is a common vocabulary (ISO 21127) for discussing information on cultural heritage and mapping it to a digital equivalent representation [Binding et al., 2008; Doerr, 2003; Eide et al., 2008; Eide & Ore, 2007; Sinclair et al., 2006; Varnienè-Janssen & Juskys, 2011]. The Functional Requirements for Bibliographic Records model (FRBR) was devised as an entity-relationship model of bibliographic data and publications [Madison, 2000; Tillett, 2004]. It documents and distinguishes the concepts forming the basis of a *Work*; the *Expression* of such *Works* in a fixed but abstract form; the *Manifestation* of such *Expressions* in physical form; and single *Items* that are exemplars of such *Manifestations*.

The CIDOC and FRBR ontologies were originally developed independently. Recognising the potential of combining these ontologies, the communities collaboratively produced FRBRoo [Doerr & LeBoeuf, 2007]. FRBRoo is the FRBR ontology expressed in an object-oriented form more compatible with that of the CIDOC-CRM, extending the CIDOC-CRM with the FRBR vocabulary. Given the relevance of CIDOC-CRM and FRBR, particularly in the repeated transmission of ideas expressed in written works, FRBRoo was the most appropriate ontology on which to base our vocabulary, suitably extended to support the description of, and the recording of relationships between, our physical and information objects. In particular, FRBRoo clarifies how the CIDOC *LinguisticObject* class and the FRBR *Expression*/*Manifestation* classes relate to each other, allowing greater clarity in representing our relationships.

Other relevant ontologies were considered:

- An extension of the CIDOC-CRM, CRM$_{dig}$, has been proposed for documenting digital objects [Doerr & Theodoridou, 2011]. While it makes significant enhancements to the CIDOC-CRM for dealing with *digital* documents, the standard form of CIDOC-CRM was more relevant for our purposes.
- We considered other ontologies documenting bibliographic resources[xiv], as well as an ontology for documenting scholarly works[xv], but they lacked sufficient depth for describing their content. The SPAR suite of ontologies for Semantic Publishing And Referencing[xvi] offer more depth and scope for detailed referencing, but the term "manuscript" in SPAR's FRBR-aligned Bibliographic Ontology (FaBiO) is explicitly used to refer to a textual work that is not 'a handwritten historical document on paper or parchment'[xvii], which is exactly the type of manuscript that we need to model.
- The OntoMedia [Lawrence, 2007] and Stories[xviii] ontologies focus on the content of a text at the expense of information about the document itself.
- SKOS could be used to represent the hierarchical structure of information content, and Dublin Core metadata provides a vocabulary for describing information about the manuscript[xix].

Each of these is relevant in part to our data; however, rather than using several ontologies representing different aspects, it was decided to adopt FRBRoo as the base ontology, borrowing terms from relevant ontologies as and where necessary, as FRBR-oo represented most aspects of the manuscript information.

Both the CIDOC-CRM model and FRBRoo have been implemented as OWL ontologies[xx] by the University of Erlangen, Germany. We use and extend these OWL implementations as extensions for the SAWS vocabulary.

## 3.2 The SAWS ontology as an extension of FRBRoo

### 3.2.1 Items of Interest

We use the term *Manuscript* to refer to the physical objects in which our texts are contained. Typically, a manuscript will contain more than just the collections of wise sayings; conversely, a collection of sayings may span several manuscripts. Consequently, in our model the fundamental unit is the *CollectionInstance*, which is an extension to the FRBRoo ontology as a combination of a *LinguisticObject* (CIDOC) and *Expression* (FRBR), corresponding to the physical instantiation of a collection of sayings in one or more manuscripts.

The other fundamental object is the *ContentItem*, which corresponds to the individual sections of interest - i.e. a saying - within a *CollectionInstance*. These may be simple assertions:

  'One cannot cover a fire with a cloak nor a shameful sin with time.'

The gnomologia also contain longer anecdotal sections:

> 'Diogenes was asked by someone why people give to beggars but not at all to philosophers, and he said, "Because, perhaps, they expect to become lame or blind but not to become philosophers."'

Here there are two components of interest: the statement itself ('Because, perhaps, they expect ... philosophers'), and a narrative text ('Diogenes was asked...'). Consequently we introduced two corresponding objects, *statement* and *narrative*.

### 3.2.2 Relationships

The definition of a vocabulary of relationships has been a key component of the research. Relationships may occur:

- within a single *CollectionInstance*
- between *CollectionInstances*
- between a *CollectionInstance* and a 'source text' (e.g. a Greek classical work or the Bible)
- between a *CollectionInstance* and subsequent texts that drew upon it.

We need a vocabulary that is not only capable of representing relationships among a specific set of texts, but is sufficiently flexible to be extended or refined to cover relationships in analogous materials. This vocabulary has been developed through collaboration between information scientists and scholars within digital humanities and manuscript studies, and is published at the permanent URL http://purl.org/saws/ontology.

Relationships that were identified include:

> Manuscript *isWrittenAt* Scriptorium
> Manuscript *isInLanguage* Language
> CollectionInstance *isWrittenBy* Scribe
> CollectionInstance *isTranslationOf* CollectionInstance
> Section *isSequentiallySimilarTo* Section[xxi]
> ContentItem *isShorterVersionOf* ContentItem
> ContentItem *isVerbatimOf* ContentItem

## 3.3 Example of the conceptual model in application

The following examples - translated into English for clarity[xxii] - illustrate how sayings develop. Item 1 is a saying attributed to Alexander the Great in a medieval Greek gnomological text, the 'Gnomologium Vaticanum'; Item 2 is an extract from Plutarch's 'Life of Alexander' (8.4.1), identified as a potential source of the saying. The text is not a direct quotation, but has been paraphrased.

1. Alexander, asked whom he loved more, Philip or Aristotle, said: 'Both equally, for one gave me the gift of life, the other taught me to live the virtuous life.'
2. Alexander admired Aristotle at the start and loved him no less, as he himself said, than his own father, since he had life through his father but the virtuous life through Aristotle.

The second example is a narrative only.

Item 3, below, is an extract from an Arabic anthology, and is attributed to Pythagoras ['Selections from the Sayings of the Four Philosophers: (B) Pythagoras' saying 18 (ed. Gutas)]. Here the source text seems to be Diogenes Laertius's 'Life of Aristotle' (5.19), shown as Item 4, although not only has the saying been translated from Greek into Arabic, it has become more pithy in translation, and the saying has been re-attributed from Aristotle to Pythagoras. Several relationship assertions may need to be used to represent the connection between the two sayings.

3. He said: "Fathers are the cause of life, but philosophers are the cause of the good life."
4. Aristotle said that educators are more to be honored than mere begetters, for the latter offer life but the former offer the good life.

## IV IMPLEMENTATION

The implementation approach has three main aspects:

- The encoding and publication of a digital archive of editions of a selected number of these texts;
- The identification and display of the links between the anthologies, their source texts, and their recipient texts;
- The building of tools to allow scholars outside the SAWS projects to link their texts to ours.

### 4.1 Encoding Individual Texts as TEI Documents

Each text is marked up in TEI XML schema developed at King's College London for the encoding of gnomologia, which is based on the TEI Manuscript schema. The structural markup reflects as closely as possible the way in which the scribe laid out the manuscript. We use the **<seg>** element[xxiii] to mark up base units of intellectual interest, such as the *statement* and its *narrative*. These base units need not have been identified as units by the scribe, but are the result of an editorial decision. Consider the case described in 4 above:

'Alexander, asked whom he loved more, Philip or Aristotle, said: "Both equally, for one gave me the gift of life, the other taught me to live the virtuous life."'

```

    
Alexander, asked whom he loved more,
Philip or Aristotle, said:
    </seg>
    
Both equally, for one gave me the gift of
life, the other taught me to live the
virtuous life.
  </seg>
</seg>
```

Each of these **\<seg\>** elements was given an **@xml:id** to provide a unique identifier that differentiates them from all other examples of **\<seg\>**:

    **\**

This provided the means to refer to a specific section of the text, and these internal identifiers were then used to generate URIs.

## 4.2 Encoding Using RDF within TEI \<relation\>

Our schema allows editors to publish the texts in accordance with TEI for Manuscripts while also supporting the identification and description of the relationships between individual units of interest. Because our data model extends the FRBRoo ontology to model the classes and relationships, a more flexible approach to encoding relationships within the TEI document is required, one which can cope with the generic and extensible nature of the ontology.

RDF triples are encoded within the TEI documents to include ontological information that is not present in the TEI markup itself, using the TEI element **\<relation\>**:[xxiv]

> **@ref** states the relationship type (from the list of relationships in the ontology);
> **@active** points to the URI of the resource that is being linked from;
> **@passive** points to the URI of the resource being linked to;
> **@resp** is used to identify the individual (or bibliographic source) responsible for asserting the relationship.

The TEI Guidelines now include the SAWS usage of the **\<relation\>** element as one of their examples:

```
<relation
  resp="http://viaf.org/viaf/44335536/"
  ref="http://purl.org/saws/ontology#isVariantOf"
  active="http://www.ancientwisdoms.ac.uk/cts/urn:cts:greekLit:tlg3017.Syno298.sawsGrc01:divedition.divsection1.o14.a107"
  passive="http://data.perseus.org/citations/urn:cts:greekLit:tlg0031.tlg002.perseus-grc1:9.35"/>
```
<div align="right">Show all  bibliography</div>

> **\<relation**
> **resp="**http://viaf.org/viaf/44335536/**"**
> **ref="**http://purl.org/saws/ontology#isVariantOf**"**
> **active="**http://www.ancientwisdoms.ac.uk/cts/urn:cts:greekLit:tlg3017.Syno298.sawsGrc01:divedition.divsection1.o14.a107**"**
> **passive="**http://data.perseus.org/citations/urn:cts:greekLit:tlg0031.tlg002.perseus-grc1:9.35**"**
> **/\>**

In other words, this example records an assertion that there is a relationship "isVariantOf", as defined by the SAWS ontology, between a passage of text in the SAWS corpus and a passage of text in the Perseus Digital Library. It also gives details of the assertion having been made by an individual who has an entry in VIAF (Virtual International Authority File), in this case Charlotte Roueché. Each value is a resolvable URI. The use of all four attributes is required in the SAWS usage of **\<relation\>**, as the project has been particularly concerned to enable users to trace responsibility for these assertions either to a specific person or to a bibliographic reference. It also serves as a means of ensuring that individual credit for making the assertions can be given.

If we want to express uncertainty about whether or not a relationship actually exists, we can use **<certainty match=".." locus="name" cert="low"/>** as a child of the **<relation>** element (where ".." points to the parent element). To express uncertainty about an attribute within **<relation>**, for instance about the type of relationship that exists, we can point to a specific attribute, e.g.: **<certainty match="../@ref" locus="value" cert="low"/>**.

## 4.3 Publishing Digital Editions

One of the main goals of the project was to enable the creation and publication of digital editions of the types of text under consideration. In designing a publication platform, it was important that we could not only present the texts and related commentaries, but also enable the user to visualise and explore the data and its relationships, and in doing this provide additional contextual information, alternative navigation options and multiple display options.

As comparisons between the texts are vital to the project, the main priority for the text display was to allow multiple texts to be viewed in parallel. Initially we used a modified version of the Versioning Machine (VM), developed at the University of Maryland (see Fig. 1).

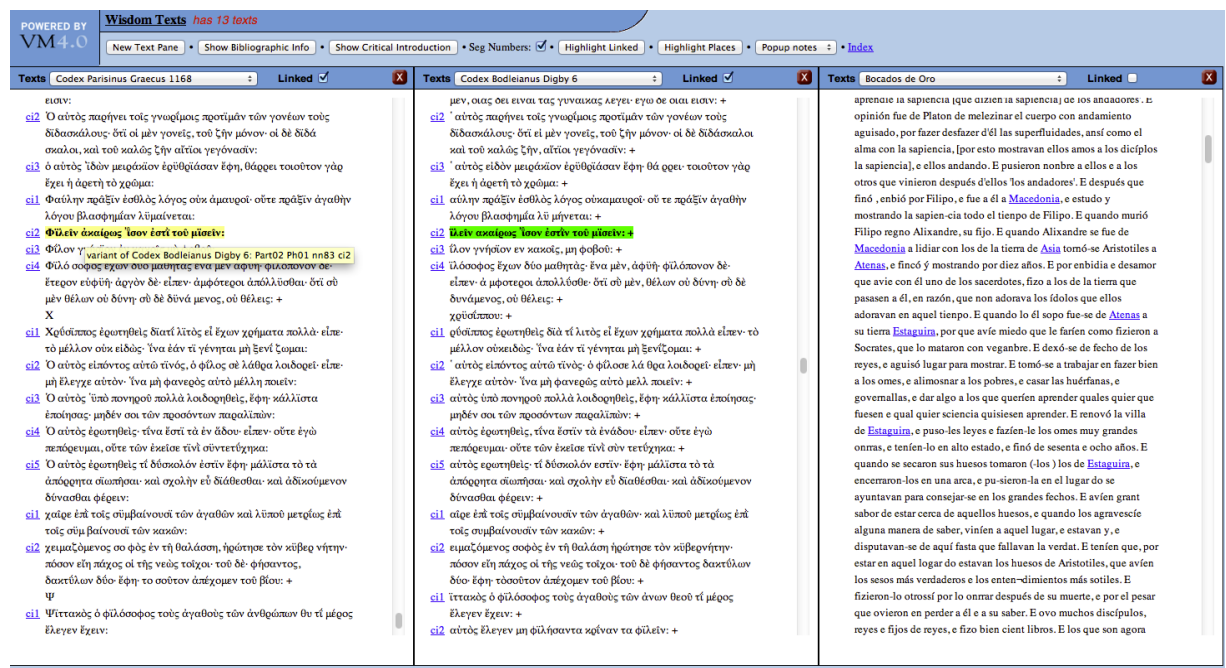

**Figure 1: Versioning Machine user interface**

As the corpus expanded and longer texts were added, we were forced by the client-side processing limitations to look for alternative systems. A number of existing parallel text display applications developed for Biblical texts were reviewed as potential candidates, but, while they showed promise, their fundamental dependency on the unique structure of Biblical books meant that they were not suitable for our purposes. Following this investigation, we developed our own parallel text viewer, which was designed to work with any text structure and to integrate the linked data and textual components of the project.

The Folioscope parallel text view is a lightweight front end integrating both the text drawn from the TEI and the relationship data provided via the Sesame triplestore. The traditional text display is provided via the Kiln framework. Developed at the Department of Digital Humanities at King's College London, Kiln uses XSLT stylesheets to create dynamic websites from XML-encoded sources. This allows for a traditional digital edition with each source and commentary presented to the reader, supported by search, indexing and contents pages. At the document (panel) level, the user is able to set display options for line and page numbers, scribal additions and identifier display, while the global settings offer multiple display options for notes, commentary, popups (for example, those produced by AWLD for Pelagios integration), and hover behavior.

Text integration is provided through direct querying of the triplestore, using display cues in combination with tooltips to indicate lines with known relations and to display the related information. Different commands align related lines in open panels or open all the related documents. Documents in the Folioscope are opening in new panels and aligned while external documents are opened in separate tabs of the browser. This allows the system to support the exploration of the texts within the library as well as those available outside it. In the figures below, the display options are identified (see Fig. 2), and examples of the resulting popups are shown (see Fig. 3).

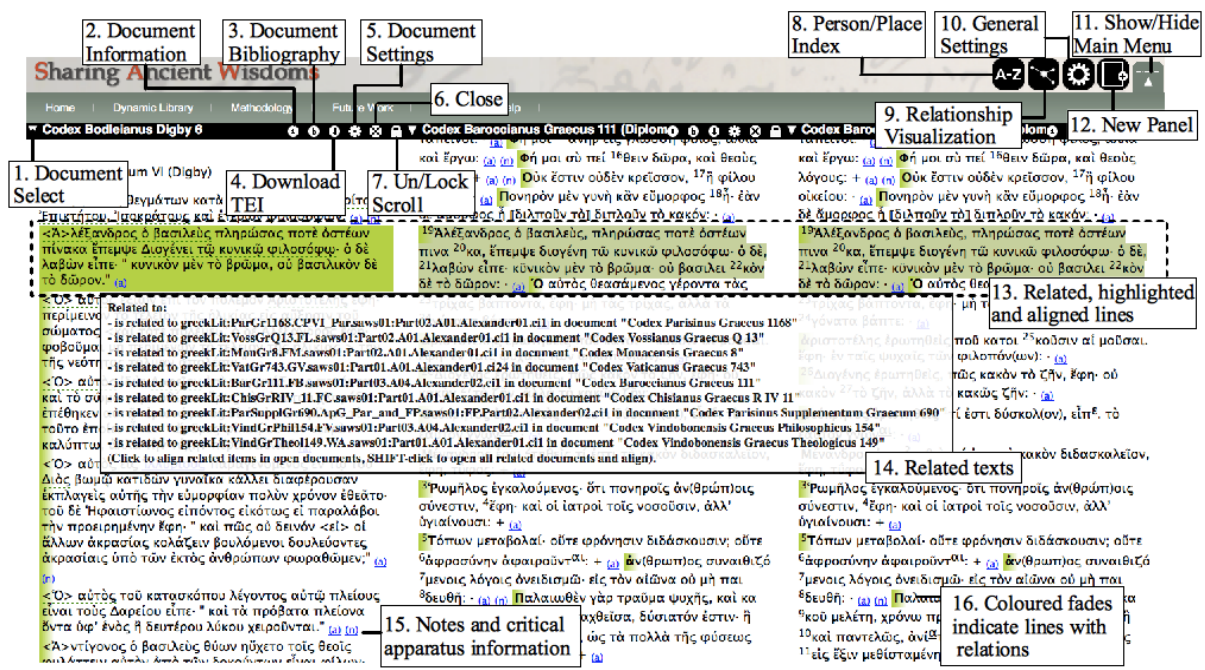

Key:
1. Drop down menu allow user to select the text to be displayed in the corresponding panel
2. Document information button - brings up a popup with information about the selected text
3. Document bibliography – brings up the bibliography of published works the corresponding text draws on
4. Option to download the underlying XML/TEI document
5. Set the line and page number display options for the corresponding panel and whether scribal additions and the link identifiers should be shown. By default line numbers and scribal additions are shown but page numbers and identifiers are hidden. Where page numbers are available they can be selected by specific publication
6. Close the corresponding panel
7. The panels can be locked to scroll in concert with the other open panels on the page or unlocked to scroll independently
8. A-Z button reveals a popup panel with the person and place indices of the document library. Available are indices for people, places (geographical reference), places (ethnic origin), personal names and ethnic group reference. Index entries display the normalized version of the term followed by a listing of each reference and the form taken in the text. Where possible links to recognized, independent, persistent identifiers are also given

9. Visualization button reveals a pop-up with a network diagram dynamically generated from the links between the selected line and any related linked data
10. Option to set the general setting for the display of popups, commentary and other notes. Different note types can be set to display as popup links (default), inline, appear in the footer when highlights or be hidden completely. Other informative popup can be set to be shown on hover (default) or be hidden
11. Option to show or hide the main menu bar with the links to the rest of the website
12. New panel button opens a new panel area in the browser
13. Clicking on a line aligns it with any related lines in open documents The selected line is highlighted, with the related lines highlighted in a complementary shade. Shift-clicking on a line opens all documents with related lines, including opening external documents in new browser windows and aligns all documents within the library
14. Hovering over a line with known relations pops up a list of all the related lines and the relationship between them
15. Links to critical apparatus commentary and other notes (see 10.)
16. Coloured fades at the beginning of a line are used to indicated that the line has related content

**Figure 2: Folioscope user interface with display options**

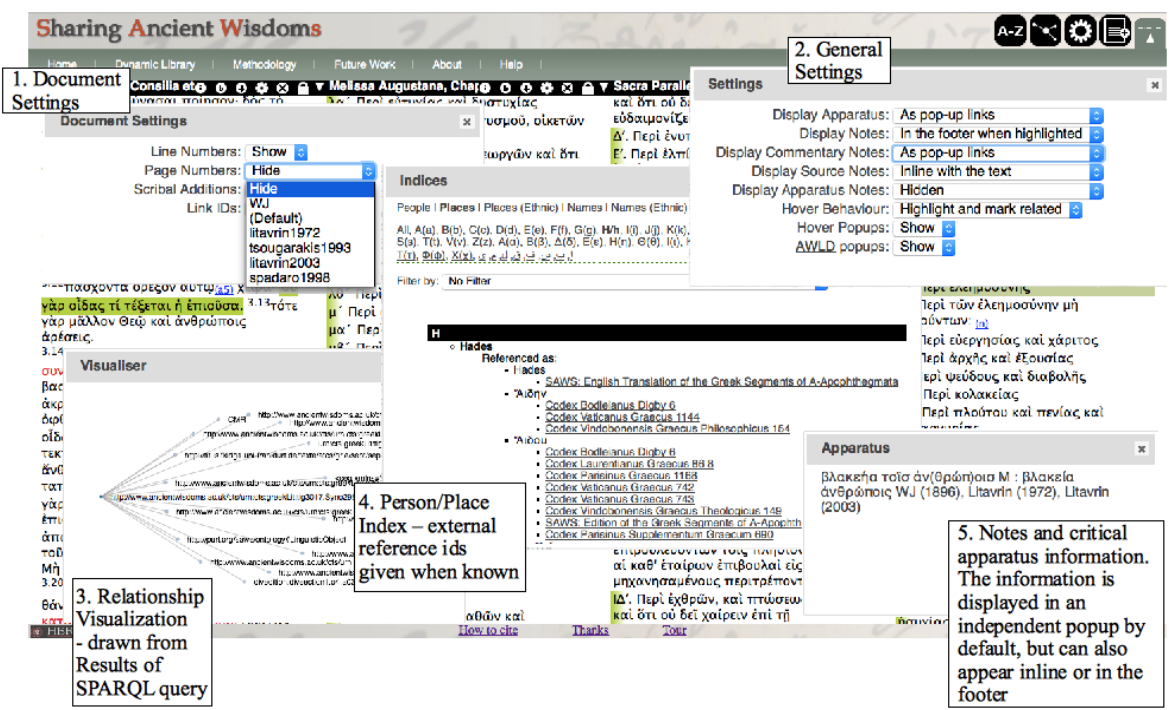

**Figure 3: Folioscope user interface with popups**

In addition to the document display options, which support scholars in refining their view of the text or comparing different editions of the same text, Folioscope also provides an automated tour of the interface, which takes the user through the options available to them, and a dynamic citation generator, which presents the referencing information for the texts currently displayed in the Folioscope viewer. These features were intended to increase the ease of use for new users, both in terms of accessing the texts in the library and in supporting the referencing of those texts in future research.

## 4.4 Linked Data

Scholars working with gnomologia also need to record links to external sources, and in particular to collections of linked data relating to the ancient world. To date, we have linked to the Pleiades historical gazetteer through references provided by the Pelagios project [Barker et al., 2012], and to documents in the Perseus Digital Library. We have also linked to information on people mentioned in a selection of the texts, through the Prosopography of the Byzantine World resource, and other resources[xxv]. We plan to extend this through participation in the Standards for Networking Ancient Prosopographies (SNAP) initiative,[xxvi] whose successful pilot project was inspired by the need for a central authority for person identifiers for the ancient world, as identified during the development of the SAWS linked dataset.

The ability to traverse links between sets of data and to discover related information serendipitously is one of the major benefits of adopting linked data for this project. It is a key part of the academic research underpinning this project and is further justified elsewhere [Solomon, 1993]. This is particularly useful where potential sources are geographically scattered, difficult to access or not widely known. As an example, the Perseus Digital Library holds a collection of Classics-related documents, which collectively contain over 68 million words, as well as an Arabic collection containing over 5 million words. Navigating such quantities of potential research material is one of the challenges faced by Classics researchers. Digitisation and cataloguing of the sources through projects like Perseus is an important step in facilitating this research, and is being enhanced further by semantic navigation such as that undertaken in by SAWS.

For example, information in the Pleiades historical gazetteer can be consulted when constructing queries. Researchers can see all texts that refer to that particular geographical location, even if different place names are used over time. This is possible because the Pleiades ontology gives a precise geographical reference for each place[xxvii]. The use of RDF and linking therefore allows us to transcend time and language boundaries to some extent[xxviii].

# V EVALUATION

## 5.1 Overall Evaluation of the Project

Throughout the life of the project, we tested our approach through invited workshops and external presentations; this enabled regular interchange with our peers. For example, we demonstrated the enhancements possible with the RDF information in a workshop in June 2012[xxix]. This highlighted several benefits, in particular the ways in which the manuscripts could be navigated. Also highlighted were ways in which the SAWS editing process could be refined, such as the editing of right-to-left directional languages such as Arabic.

This demonstration prompted useful constructive feedback, leading to the identification of further relationship types. It also prompted scholarly debates following the identification of different interpretations of the notion of translation (generated by the requirement to formalise collaboratively their tacit knowledge). Ongoing consultation with manuscript scholars provided formative evaluative feedback for further developments. These included technical collaborations with the Islandora team in Prince Edward Island, Canada, with whom we developed a TEI to RDF mapping for automatic extraction of the RDF triples inherent in the TEI markup and the triples encoded in **<relation>** elements [Jordanous, Stanley & Tupman 2012; Tupman, Jordanous & Stanley 2013]. This mapping was also deployed in the Islandora Critical Editions Solution Pack, a repository-based software tool for managing digital editions produced by the Editing Modernism in Canada (EMIC) project in conjunction with the Canadian company Discovery Garden. The work was implemented using XSLT, forming a basis for further mappings and transformations. In particular, we explored how the Dublin Core metadata model and the FRBR-oo ontology can be used to enhance the TEI to RDF mappings, for a semantically rich vocabulary.

Overall, the project successfully achieved:

- Digital edition of manuscripts published using TEI and RDF annotations.
- Manuscripts to be navigable through structural and semantic links.
- Semantic content in manuscripts to be searchable and queryable through extraction of RDF information.
- Positive impact within the philological community, particularly those researching medieval manuscripts

## 5.2 Evaluation of the SAWS Ontology

The evaluation of the SAWS ontology had two aspects: (i) demonstrating the validity (logical consistency) of the ontology, and (ii) assessing the ability of the ontology to express what we wanted to express, that is its fitness for purpose.

The validity of the ontology was checked using reasoners built in to the Protégé tool, which was used to develop the ontology. Reasoners check that logical statements within the ontology are consistent, with no contradictions. Application of the reasoners within Protégé highlighted no such inconsistencies in the SAWS ontology, and thus demonstrated the validity of the ontology.

Brank et al. (2005, p. 1) outline four different approaches to evaluating how useful and representative an ontology is:

- 'those based on comparing the ontology to a "golden standard" (which may itself be an ontology …);
- those based on using the ontology in an application and evaluating the results …;
- those involving comparisons with a source of data (e.g. a collection of documents) about the domain to be covered by the ontology …;
- those where evaluation is done by humans who try to assess how well the ontology meets a set of predefined criteria, standards, requirements, etc.'

We have adopted those approaches relevant to our ontology usage to evaluate the quality of the SAWS ontology. The first evaluative method identified by Brank et al. (2005) is to compare the ontology to an existing 'golden standard'. In our case, we determined that the best candidate for an existing 'golden standard' ontology for recording information about documents is FRBRoo (based on CIDOC-CRM and FRBR). As described in Section 3.1, it is considerably more difficult to use the other mentioned ontologies to express our data to the required level of detail, due to the lack of available vocabulary. Some significant types of data cannot be expressed using the alternatives to FRBRoo, hindering us from carrying out a meaningful comparison of them in relation to the SAWS ontology. We thus concluded that FRBRoo is the best candidate for a 'golden standard'.

The requirements for an ontology that were identified from the collaboration between domain experts and technical observers during the SAWS project were successfully mapped onto the existing FRBRoo ontology, which has undergone extensive review from both the CIDOC and FRBR communities (recall that FRBRoo is a 'harmonisation' of the CIDOC CRM cultural heritage model and the FRBR model for bibliographic records), as well as by users of FRBRoo. After this mapping process, the FRBRoo ontology could be used to express some of the required data concerning relationships between objects; however, FRBRoo was deemed deficient for our purposes, as its terms did not allow for the required level of detail to be expressed in the data.

Our work extending the FRBRoo ontology has aimed to fix these deficiencies, so that the SAWS ontology provides a level of granularity sufficient for scholars working with gnomologia to be able to represent their scholarly knowledge at a desired level of detail (this extension work is described in Section 3.2). The deficiencies can be seen most clearly by examining where new terms have had to be introduced. The most significant deficiencies are illustrated in Fig. 4, which is taken from screenshots of the ontology as viewed in Protégé . The darker circles indicate existing terms in CIDOC-CRM or FRBR; the precise origin is indicated by the blue dotted circles in the prefix of each term, before the colon, but they are all included in the FRBRoo harmonisation of CIDOC-CRM and FRBR. The lighter circles indicate SAWS ontology terms, which we have added to increase the level of expressiveness of the ontology to the level desired by the domain experts. Key elements that required more detail than provided by FRBRoo were Linguistic_Object and Person (both highlighted in red dotted rectangles in the figure). To allow domain experts to record their scholarly expertise more comprehensively, the SAWS ontology needed terms representing more fine-grained information about the types of people interacting with the manuscripts and other textual documents under investigation. It is important to be able to record whether a given 'Person' was a manuscript scribe, or an author to whom a saying was attributed, or an editor of a later edition of sayings, for example. These people play entirely different roles as they interact with the manuscripts, and information about their role contributes to our understanding of the manuscripts, their content and their transmission. We also needed to be able to categorise types of textual documents in more detail than a single generic type of 'Linguistic_Object'; for example, is this 'Linguistic_Object' a compilation of sayings within the manuscript, or a later edited collection? Introducing extra terms in the SAWS ontology provides us with more vocabulary with which to express more detailed relevant scholarly knowledge.

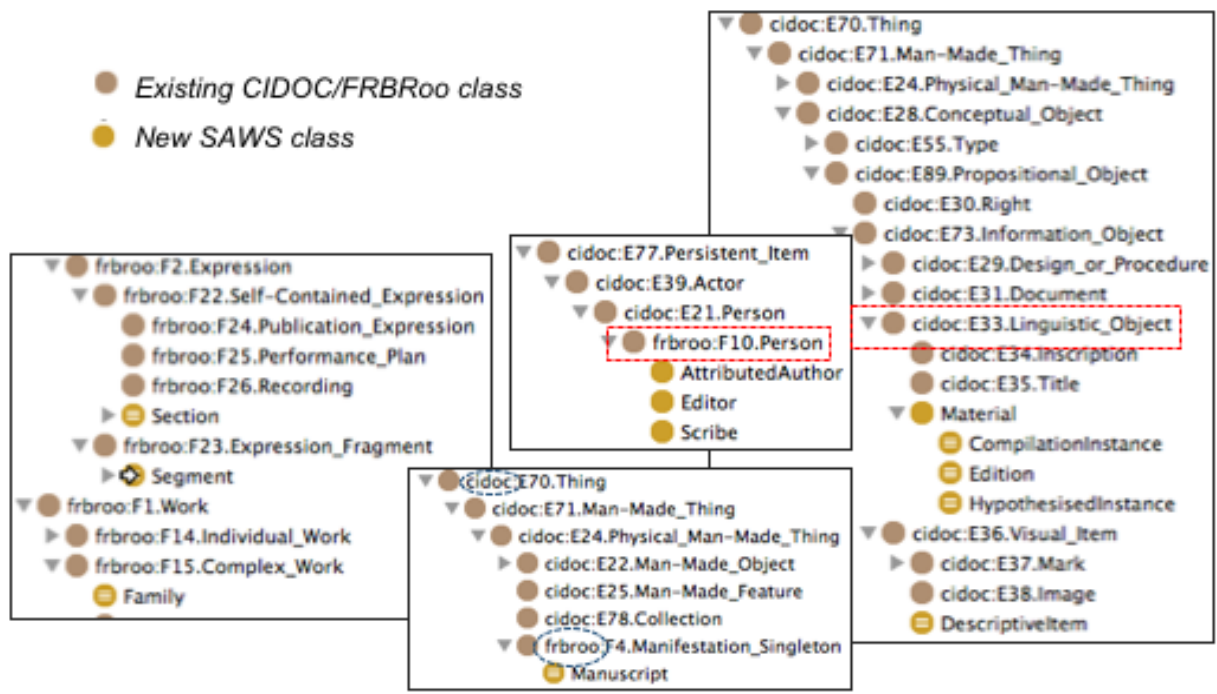

**Figure 4: Comparison between CIDOC/FRBRoo and the SAWS ontology**

Our mapping and extension work for the SAWS ontology has been recognised as a CIDOC-CRM-compliant extension [Alexiev et al., 2013, p. iii]. This means that the SAWS ontology extends the underlying FRBRoo (CIDOC+FRBR) ontology in a manner that is officially accepted by the community of users and validators for the FRBRoo ontology and the original CIDOC-CRM model.

How do we know that the resulting SAWS ontology has indeed dealt with the deficiencies of the FRBRoo ontology to an acceptable level of detail? To answer this, the resulting ontology was presented to the domain experts in the form of a vocabulary they could use to express relationships between and within the manuscripts they studied. Formative feedback solicited from the domain experts at various stages was used to refine the ontology further, for another stage of evaluation performed by people (usually domain experts) who 'try to assess how well the ontology meets a set of predefined criteria, standards, requirements, etc.' [Brank et al., 2005, p.1]. Feedback solicitations occurred on an ongoing basis with domain experts within the SAWS team, including a number of meetings specifically dedicated to evaluating the ontology as it was at that time. These meetings included approximately one face-to-face meeting of the SAWS team each 6-12 months, with more frequent online correspondence taking place between face-to-face meetings; the SAWS team represented three groups of domain experts, each consisting of a lead researcher and at least one other researcher or PhD student researcher. We also conducted a number of workshops with domain experts outside the SAWS team. Eleven workshops were run during the lifetime of the project, with between 5 and 30 participants, in locations across Europe such as Vienna, London, Göttingen and Lund, as well as further afield in the USA (see http://www.ancientwisdoms.ac.uk/about/workshops/ for further details.) The workshops were complemented by feedback received during conference presentations of the work given by members of the SAWS team (see http://www.ancientwisdoms.ac.uk/about/presentations/ for further details). A fuller list of people and groups that we interacted with during the SAWS project can be found in mpa format at: https://www.google.com/maps/d/viewer?ll=44.902577104648564%2C4.927368999999999&spn=35.093525%2C233.668213&hl=en&t=h&msa=0&source=embed&ie=UTF8&mid=1b6XicX9DLfR1blTLpdhueydqfy0&z=2 .

A number of modifications were made to the ontology as a result of these feedback sessions and workshops, many of them involving making the wording of definitions or terms better reflect their shared understanding of the terminology to be represented in the ontology. One example of a more impactful and recurring set of revisions was mentioned earlier in Section 5.1, namely the notion of translation. As noted in the earlier section, ongoing discussion of the appropriateness and accuracy of the terms revealed an interesting dichotomy in the tacit understanding of the word 'translation'. Some domain experts focused their study of translation on the source and the end result, and within this interpretation there was also some variation in the level of attention paid to the source document(s). Other domain experts focused on the process of translation, with more interest in how the translation was done. This conceptual variance in the interpretation of translation was only highlighted during the evaluative discussions, and led to some fascinating realisations between the experts about their tacit, unacknowledged differences in emphasis. This was a nice example of how the process of refining and improving a model of knowledge can also feed back to an improved *shared* understanding of the terms. The hierarchical set of terms for translation used by SAWS was developed and refined over time, to be comprehensive enough to cover these differing interpretations of translation, so that all the domain experts present during these discussions expressed satisfaction with the final result; they were able to express translation-based relationships between collections of sayings without compromising their own personal view of translation.

Another introduction made to the SAWS vocabulary, as a result of one of the external workshops, was the term 'Hypothesised_Instance' (see Fig. 4). 'Hypothesised_Instance' refers to a collection or compilation of sayings for which there is no physical evidence, but which we believe existed and which we want to be able to talk about as part of our research. Several domain experts highlighted the need for this term, for example to discuss the hypothetical existence of an earlier collection which is now lost but which, it is hypothesised, was an important source material for several later manuscripts.

Typically, in the early stages of the project the ontology underwent significant revision during these evaluation sessions. It is difficult to claim that any ontology can reach a stage of being finished and complete; knowledge develops quickly and terms within a discipline become used in different ways over time. Certainly with the SAWS ontology there are still a small number of terms that are not universally agreed upon by all domain experts involved. However some discrepancies were to be expected, as we are not dealing with objective facts but with scholarly opinions, interpretations and assertions, which are open to debate. For example, as mentioned above, evaluation of the ontology helped the domain experts to identify different interpretations of translation of which they had previously been unaware. Overall, as the evaluations towards the end of the project resulted in a much-reduced number of revisions from internal and external experts, we are confident that the SAWS ontology has reached a state of some stability, and has met with the approval of many domain experts. As will be discussed in the next section, it has already been adopted for reuse by a number of other teams working in this area.

One interesting observation from external domain experts was that the ontology was too focussed towards medieval anthologies of wise sayings. The aim of the SAWS ontology was in fact to represent this type of text; however, our use of the FRBRoo as the underlying ontology helps to provide ways of dealing with other types of texts. This issue could be addressed by improving the presentation of the ontology (e.g. by including more of the underlying FRBRoo ontology) rather than by improving the ontology itself. This would make the SAWS ontology more widely applicable in closely related domains, without losing its original focus and scope.

The evaluation of the overall project approach contributes to another method identified by Brank et al. (2005) for evaluating the quality of an ontology: to use the ontology within an application, and to evaluate the results. In the next section we highlight and focus on this particular type of evaluation, which we see as the critical part of assessing this work.

### 5.3 Evaluation of the Digital Approach of SAWS

As noted above, one way of evaluating the success of an ontology-based approach to representing information is to evaluate how the ontology is used in an application, to assess the usefulness and accuracy of the results. We adopted this approach as part of the ongoing evaluation throughout the lifetime of the SAWS project, as a feedback mechanism to continually develop the ontological work [for example, see the mid-project comments on evaluation in Jordanous et al., 2012]. As reported above, the ontology was refined at various stages throughout the project. Critically, the resulting ontology is successful to the degree that it allows domain experts to record their tacit knowledge and expertise in digital form. Through evaluation of the use of the SAWS ontology in application, the following feedback on the usefulness of our digital approach was identified:

- Through marking up the manuscripts in TEI XML we have made these collections of sayings available in digital form with structured content, removing the accessibility problems to the original physical manuscripts. The text of the manuscripts has been supplemented with expert knowledge, much as would happen when producing a critical edition.
- The mark-up process has been undertaken both by experts in this area and non-experts supervised by experts and given brief training. Especially for larger-scale markup projects, the markup process can be time-consuming and it is useful to be able to share

                    

this workload without needing to recruit several people with detailed expert knowledge. The process of tagging the digital versions of the collections is modular and can be performed in a distributed way, across a number of people, with the experts being able to add more detail from their specialist knowledge whilst sharing the more repetitive markup with others.

- The markup provided through the SAWS TEI schema caters for different stages of annotation: from quick annotations 'out in the field', when researchers are actually at the physical location where the manuscript is kept; initial editing of structure and brief observations; detailed analysis; through to the publication of a critical edition of the manuscript. This models the analytical processes and stages that such researchers are already familiar with in their work.

- There is a growing desire to make more of the XML documents by including relationship information within the TEI markup itself, to which the SAWS approach has made a contribution. Information on how documents are related and how links exist within documents is extracted from the analysis to be included in the TEI editions of the manuscripts. This enhances the semantic content of these electronic versions. Highlighting the contribution of the SAWS project to these developments: example markup from SAWS TEI documents has been included in the official TEI guidelines documentation, acting as an example of best practice to follow in incorporating RDF within TEI.

- Further evaluation of the SAWS ontology and the overall approach will be seen over time, through the use of the SAWS ontology in other digital humanities applications. To date, we have collaborated with the *Corpus der arabischen und syrischen Gnomologien* (CASG) project, Halle, Germany to help them adopt a SAWS approach (see xxx, paper submitted to this special issue by CASG). We have also worked with the *Monastic Paideia* (MOPAI) project in Lund, Sweden, to advise on how they can use SAWS [Johnsson & Åhlfeldt, 2015]. From citations we see that SAWS has directly inspired: the approach taken by the BIBLIMOS project for publishing ancient scientific Mauritanian manuscripts [Markhoff et al., 2015]; the use of the FRBRoo ontology as a base for the OMOS (Ontologie sur des Manuscrits Ouest Sahariens) ontology for Western Saharan Manuscripts [Diakite and Markhoff, 2015]; an ontology representing data about proverbs [Zhitomirsky-Geffet et al., 2015]; and a semantic study of Dante Alighieri's philosophical essay 'Convivio' [Bartalesi et al., 2013]. SAWS has also influenced the use of RDF within TEI markup in the BIA-NET project, for archiving Ancient Roman Law texts [Spampinato & Zangara, 2013] an an ontology for Sumerian literary narratives (which also extends FRBRoo) [Nurmikko-Fuller, 2015].

## VI FUTURE WORK

The SAWS project has received many indications of interest from the philological community. On a longer-term basis, the success of the SAWS approach will be demonstrated by the future and ongoing adoption of a SAWS-style approach by others across this community, for editing, annotation and publishing of digital manuscript editions. A particular marker for success will be the linking to and from SAWS manuscripts by scholars outside of the SAWS research team, particularly if the SAWS digital editions become the canonical reference point for the manuscripts digitised during the project. Another indicator is the possibility of the SAWS approach being adopted by researchers outside the immediate target audience of manuscript scholars, for example those studying modern texts or other objects represented in TEI, for example the MEI (Music Encoding Initiative) community [Roland, 2002].

The SAWS approach allows us to extract triples from the marked up TEI documents, to be stored in a triple store and queried with SPARQL. With the data in a queryable form, this opens up a whole host of exciting possibilities. The primary aim is to enable the creation of digital analysis and information extraction tools for the immediate target audience (digital humanities researchers), to collect information. Outside the immediate audience, data on wise sayings and how they have evolved over transcription and transmission would also be of interest to linguists, social scientists and historians. The collections of sayings could also be exploited for potential 'pop'-applications outside the academic sphere of interest, such as online or mobile apps to generate wise sayings in appropriate contexts.

In SAWS we have created a framework for others to use and extend; a growing network of interconnected information. As the body of material of interest in this field is potentially very large, we do not view the project as creating just a digital, online edition, although this will be one result of the project, but rather as creating the kernel for a much larger corpus of interrelated digital editions. We envisage this as a SAWS `hub' for enabling related projects to annotate and link their own texts. The research value of such a corpus would be much greater than the sum of its parts, and would increase dramatically once a `critical mass' was reached.

Many of the subsequent contributions to this corpus will, of course, be carried out by other researchers; as described above, we worked closely and creatively with, the Monastic Paideia project in Lund, Sweden, and the CASG project in Halle Germany (for which see the paper by the Halle team in this journal special issue), training researchers from both projects. If such undertakings are to be able to interoperate and contribute to the wider corpus, rather than existing as a collection of separate editions (which would be of much less value to researchers), it is important that everyone 'speaks the same language' regarding how this material is represented in digital form, both semantically and technically; herein lies a significant part of our long-term provision for future scholarship. Moreover, these contributions are likely to be made over a long duration, certainly long in relation to the speed of technical developments, so our approach must be such as to allow migration, without loss of information, as the technological environment changes; our adoption of current standards and reusable ontologies assists this aim.

SAWS has also inspired development outside of philological research. The integration of externally linked information popups showed a clear gap in the (digital) publication. While classical geography is well supported by the Pelagios project, the initial plan to extend the AWLD pop-up library to other sites and provide complementary information about persons mentioned in the text was limited by the lack of suitable sources to which we could link. The lack of a single resource for unique canonical identifiers, such as that which Pleiades offers for places, for classical persons is now being addressed by the SNAP:DRGN project as a direct result of the SAWS initiative.

The development of user-facing tools is another area in which it will be essential to work closely with the scholars who will (or who may) use the tools. We cannot assume that the users will be au fait with the technology, neither can we assume that all scholars will have access to specialists in this area, so the tools must be usable with the help only of standard on-line help and documentation; releasing the tools for reuse by other scholars is important.

Further work is required in the community to ensure that the tools are generalised and enabled to deal with a wide variety of textual structures.[xxx]

## VII CONCLUSIONS

The work described in this paper allows us to exploit semantic web technologies for a better understanding of the highly interconnected medieval manuscripts known as gnomologia (collections of 'wise sayings'), to obtain a greater understanding of the cultural dynamics of the medieval Mediterranean world.

We achieved this through:
- The publication of the texts as TEI documents, with embedded RDF to record relationships identified by the editors;
- A framework allowing the identification of sections of intellectual interest within texts, relationships between texts, and the recording of these as annotations;
- Links internally within/between documents and externally to relevant sources of linked data outside our collections;
- A methodology that can be used by other scholars to analyse and publish analogous material.

We are thus publishing not only digital editions, but also relationships and semantic annotations within and between those texts, creating a network of relationships and providing a framework upon which others can build. Ultimately we will produce a network of digital editions of these manuscripts, enhanced by a network of semantic annotations and relationships.

We advocate a methodology for making these manuscripts accessible in a way not previously possible, together with tools to support the researcher in studying the collections. By using these textual relationships to analyse the flow of knowledge between texts and cultures, SAWS will enable a better understanding of the processes of cultural exchange between civilisations, and in particular of the cultural dynamics across the centuries of Greek and Arabic thought.

## VII ACKNOWLEDGEMENTS

The SAWS team comprised sub-teams from King's College London, UK; Newman Institute, Uppsala; and the University of Vienna. The project was funded by HERA (Humanities in the European Research Area).

---

i Examples of such publications include: 'Inscriptions of Aphrodisias', http://insaph.kcl.ac.uk/iaph2007/ ; 'TheArchimedes Palimpsest Project', http://archimedespalimpsest.net/ ; 'Jane Austen's Fiction Manuscripts', http://www.janeausten.ac.uk ; 'Biblioteca Virtual Miguel de Cervantes', http://www.cervantesvirtual.com/ ; 'Perseus Digital Library', http://www.perseus.tufts.edu/. All last accessed October 2012.
ii http://www.ancientwisdoms.ac.uk
iii Their significance is underlined by the fact that Caxton's first imprint (the first book ever published in England) was one such collection [Caxton, 1877)].
iv A concept not always without difficulties, but these can be ignored for the purposes of this discussion.
v They concluded that while 'text is best represented as an ordered hierarchy of content object... the hierarchical model can allow future use and reuse of the document as a database, hypertext, or network' (DeRose et al., 1990, p. 3); 'Some structure cannot be fully described even with multiple hierarchies, but requires arbitrary network structures' (DeRose et al., 1990, p. 19).
vi RDF (2004). 'RDF/XML Syntax Specification (Revised)'. http://www.w3.org/TR/rdf−syntax−grammar/ . Last accessed October 2012.
vii Also 'Text Encoding Initiative / Tracker / Feature Requests / Encoding RDF relationships in TEI - ID: 3309894 (Discussion)'. http://sourceforge.net/tracker/?func=detail&atid=644065&aid=3309894&group_id=106328 . Last accessed October 2012.
viii 'OntoMedia Ontology - Working Model', http://www.contextus.net/ontomedia. Last accessed October 2012.
ix http://www.tei-c.org/SIG/Ontologies/
x As reported in the SIG's meeting minutes at http://wiki.tei-c.org/index.php/SIG:Ontologies
xi As acknowledged in the documentation at http://www.tei-c.org/SIG/Ontologies/guidelines/guidelinesTeiMappableCrm.xml
xii See discussion at http://sourceforge.net/tracker/?func=detail&atid=644065&aid=3309894&group_id=106328, and guidelines at http://www.tei-c.org/release/doc/tei-p5-doc/en/html/ref-relation.html
xiii 'CIDOC-CRM', http://www.cidoc-crm.org
xiv 'Bibliographic Ontology', http://purl.org/ontology/bibo/ ; 'Simplified Ontology for Bibliographic Resources (SOBR)', https://gist.github.com/1331983 . Both last accessed October 2012.
xv 'Scholarly Works Application Profile', http://www.ukoln.ac.uk/repositories/digirep/index/Scholarly_Works_Application_Profile. Last accessed October 2012.
xvi 'SPAR', http://purl.org/spar . Last accessed October 2012.
xvii In the FaBiO documentation for 'fabio:manuscript', http://www.essepuntato.it/lode/http://purl.org/spar/fabio#d4e3689 . Last accessed October 2012.
xviii M. O. Jewell, et al. (2010). 'Stories Ontology - Working Draft'. http://contextus.net/stories/ . Last accessed October 2012.
xix 'SKOS Simple Knowledge Organization System Reference', http://www.w3.org/TR/skos-reference/ ; 'DCMI Metadata Terms', http://www.dublincore.org/documents/dcmi−terms/. Both last accessed October 2012.
xx 'eCRM', http://erlangen-crm.org/current/ ; 'EFRBRoo', http://erlangen-crm.org/efrbroo/ . Both last accessed October 2012
xxi A Section is a structural part of a CollectionInstance. One Section has a slightly different sequence to another Section but is related, e.g. through editorial decisions made whilst copying.
xxii By the project team.
xxiii http://www.tei-c.org/release/doc/tei-p5-doc/en/html/ref-seg.html
xxiv http://www.tei-c.org/release/doc/tei-p5-doc/en/html/ref-relation.html. See also 'TEI Ontologies SIG'. http://www.tei−c.org/Activities/SIG/Ontologies/ and 'Text Encoding Initiative / Tracker / Feature Requests / Encoding RDF relationships in TEI - ID: 3309894 (Discussion)'. http://sourceforge.net/tracker/?func=detail&atid=644065&aid=3309894&group_id=106328 . Both last accessed October 2012.
xxv http://www.pbw.kcl.ac.uk/
xxvi https://snapdrgn.net/

xxvii For example the place "Aphrodisias" (URI http://pleiades.stoa.org/places/638753) was known by the names Ninoe (in the Classical period); Aphrodeisias (Hellenistic-republican, Roman periods); Lelegon polis (unspecified period); Stauropolis (Late-antique period); and Aphrodisias (Roman, Late-antique periods), but these placenames are linked together by Pleiades. Pleiades also disambiguates between Aphrodisias located in modern-day Turkey and the Aphrodisias in modern-day Spain (URI http://pleiades.stoa.org/places/255978/).

xxviii E.g. the person "Aristotle" can be represented by the URI http://dbpedia.org/resource/Aristotle independently of whether he is referred to as Aristotle, Ἀριστοτέλης, أرسطو, Aristoteles, or Aristóteles.

xxix Demo available at http://www.ancientwisdoms.ac.uk/media/data/texts.html.

xxx A Django project - with one application serving the Folioscope app and one serving and managing the data - is under development which will allow Folioscope to be deployed easily as part of a Django-based site.